\documentclass{article}

\usepackage{arxiv}

\usepackage[utf8]{inputenc} 
\usepackage[T1]{fontenc}    
\usepackage{hyperref}       
\usepackage{url}            
\usepackage{booktabs}       
\usepackage{amsfonts}       
\usepackage{nicefrac}       
\usepackage{microtype}      
\usepackage{lipsum}
\usepackage{graphicx}
\graphicspath{ {./images/} }

\title{Investigating Moral Foundations from Web Trending Topics}

\author{
 Jean Marie Tshimula \\
  Department of Computer Science\\
  Université de Sherbrooke\\
  QC J1K 2R1, Canada \\
  \texttt{kabj2801@usherbrooke.ca} \\
   \And
 Belkacem Chikhaoui \\
  LICEF Research Center\\
  Université TÉLUQ\\
  QC H2S 3L5, Canada\\\
  \texttt{belkacem.chikhaoui@teluq.ca} \\
  \And
 Shengrui Wang \\
  Department of Computer Science\\
  Université de Sherbrooke\\
  QC J1K 2R1, Canada \\
  \texttt{shengrui.wang@usherbrooke.ca} \\
}

\begin{document}
\maketitle

\begin{abstract}
Moral foundations theory helps understand differences in morality across cultures. In this paper, we propose a model to predict moral foundations (MF) from social media trending topics. We also investigate whether differences in MF influence emotional traits. Our results are promising and leave room for future research avenues.
\end{abstract}

\section{Introduction}
Nowadays, media, governments, and organizations keep a constant eye on social media for uncovering the very latest trends. Web trending topics include various opinions on the matters covered in the community. The exploitation of trending topics depends strongly on the goals and interests of each entity. For instance, healthcare organizations may study social media language associated with trending topics to track mental health, symptom mentions and changes in people's well-being throughout the pandemic. A government may continuously monitor web trends for identifying rapidly growing divisive and controversial topics that can produce political tensions and an endlessly chaotic situation, and tarnish the country's reputation and image. 

While some trends may involve strong emotional and passionate debate, we need to be clear about the fact that some people may be influenced by a wide variety of social and emotional forces and even indulged in emotional manipulation. Research shows that psycholinguistic features and sentiment analysis can provide substantial insights into the issues that affect emotional stability, intensity, and reactions \cite{Kahn2007}. In this paper, we are interested in examining moral foundations theory to discern moral differences in a broad spectrum of opinion on social media. It is of great importance to understand moral differences at cultural and individual levels because morality guides human social interactions and can potentially conduct to divergence of opinion, polarity, violence, and hostility when there are moral shocks within a community. Understanding moral foundations can yield promising results in terms of perceiving the intended meaning of the text data because the concept of morality provides additional information on the unobservable characteristics of information processing and non-conscious cognitive processes \cite{graham2014}, \cite{ece2020}. Researchers believe that five psychological dimensions are functioning as foundations for moralities around the world \cite{graham2007}. 

The five moral foundations are care/harm, fairness/cheating, loyalty/betrayal, authority/subversion, and purity/degradation. Each dimension possesses virtues and vices: 

\begin{itemize}
    \item Care/harm is associated with the protection of self and others from harm's way. 
    \item Fairness/cheating is related to the evolutionary process of reciprocal altruism and generates ideas of justice, rights, and autonomy.
    \item Loyalty/betrayal is based on the expressions of self-sacrifice for both ends of the virtue-vice spectrum, such as patriotism-betrayal, faithfulness-unfaithfulness.
    \item Authority/subversion underlies virtues of leadership and followership, including deference to legitimate authority and respect for traditions.
    \item Purity/degradation is associated with sanctity in the virtue dimension and degradation and pollution in the vice dimension.
\end{itemize}

To the best of our knowledge, our paper is the first to address the problem of moral foundations and emotional traits in web trending topics. We utilized Moral Foundations Dictionary (MFD)\footnote{https://moralfoundations.org} and MoralStrength \cite{Araque2020}\footnote{https://github.com/oaraque/moral-foundations} in which the five aforementioned psychological dimensions are considered. We combined MFD and MoralStrength and removed duplicate words for each dimension. We considered the words of each dimension as labels and then resorted to zero-shot text classification (ZSC) \cite{yin2019} to assign a probability to each label in text data. 

With the probabilities of labels obtained, we conducted 10-fold cross-validation to split our training and testing sets and trained a support vector machine (SVM) classifier to predict each dimension. Additionally, we used MFD-MoralStrength together with the Linguistic Inquiry and Word Count (LIWC) to investigate whether differences in moral foundations influence some emotional traits. LIWC dictionary is a widely used psychometrically validated system for psychology-related analysis of language and word classification \cite{Pennebaker2015}. LIWC includes word categories that have pre-labeled meanings created by psychologists. LIWC categories have been also been independently evaluated for their correlation with psychological concepts. For each input sequence, we computed the number of observed words, using LIWC and focusing on five LIWC subcategories of psychological processes: positive emotion, negative emotion, anger, anxiety, and sadness. Specifically, we performed the Pearson correlation ($\it r$) between the MFD-MoralStrength word scores and LIWC features extracted from the text data.

We conducted extensive experiments to investigate two trending topics: coronavirus lockdown measures and the Western Exit (WEXIT) separatist movement in Canada. We collected 857,294 coronavirus lockdown-related tweets posted between 12 March 2020 and 25 May 2020. Specifically, we extracted tweets bearing the words or hashtags: covid, coronavirus, \#StayAtHome, or \#StayHome. For the WEXIT movement, we collected nearly 78,000 tweets, posted between 19 Oct. 2019 and 3 March 2020, using the hashtags \#wexit and \#wexitnow. To preprocess the data, we limited our set to Canada geolocated tweets and removed tweets written in a language other than English or French.

\section{Discussion}

Tables \ref{tab3EWederf234ERR}\ and \ref{tab3erws2eddd1} show the results of the Pearson correlation ($r$) between moral word scores and psycholinguistic features extracted from the lockdown and WEXIT dataset, respectively. For coronavirus lockdown, we observe that all correlations between {\tt positive emotion} and virtue moral foundations ({\tt Care}, {\tt Fairness}, {\tt Loyalty}, {\tt Authority}, {\tt Purity}) as well as all correlations between {\tt negative emotion} and vice moral foundations ({\tt Harm}, {\tt Cheating}, {\tt Betrayal}, {\tt Degradation}) are statistically significant at $p<0.001$; except for one vice moral foundation, {\tt Subversion} ($p>0.05$). We report that {\tt Degradation} is associated with a relatively high correlation with {\tt negative emotion}, {\tt anger}, {\tt anxiety}, and {\tt sadness} ($p<0.001$) during the lockdown period. This reflects the feelings that touch directly on mental health. 

\begin{table*}[!ht]
\setlength\tabcolsep{1.5pt}
\centering
\caption{Pearson correlation ($\it r$) between moral word scores and LIWC features during coronavirus lockdown in Canada.}\label{tab3EWederf234ERR}
\begin{tabular}{lcccccccccc} \toprule
            & Care & Harm & Fairness & Cheating & Loyalty & Betrayal & Authority & Subversion & Purity & Degradation \\ \toprule
Pos. emotion & 0.151  & 0.014  & 0.183  & 0.161 & 0.246 & -0.021 &  0.053 & 0.027 & 0.186 & -0.084\\

Neg. emotion & -0.093 & 0.131 &  0.054 & -0.022 & -0.042 & -0.030 & 0.017 & 0.235 & -0.094 & 0.191\\

Anger & -0.041 & 0.128 & 0.105 & -0.019 & -0.093 & 0.191 & -0.052 & 0.176 & 0.058 & 0.173 \\

Anxiety & 0.038 & -0.189 & 0.033 & 0.101 & 0.097 & -0.013 & -0.006  & 0.238 & 0.003 & 0.219 \\

Sadness & 0.195 & 0.081 & 0.132 & 0.004 & -0.093 & 0.155 & 0.032 & 0.107 & -0.031 & 0.112 \\

\toprule
\end{tabular}

\end{table*}

\begin{table*}[]
\setlength\tabcolsep{1.5pt}
\centering
\caption{Pearson correlation ($\it r$) between moral word scores and LIWC features during a specific period of WEXIT.}\label{tab3erws2eddd1}
\begin{tabular}{lcccccccccc} \toprule
            & Care & Harm & Fairness & Cheating & Loyalty & Betrayal & Authority & Subversion & Purity & Degradation \\ \toprule

Pos. emotion & 0.252 & -0.044 & 0.209 & -0.093 & 0.102 & -0.097 & 0.160 & 0.122  & 0.103 & 0.017 \\

Neg. emotion & 0.057 & 0.203 & 0.073 & 0.154 & -0.046 & 0.024 & 0.188 & 0.001 & 0.020 & 0.113 \\

Anger & -0.026 & 0.145 & 0.003 & -0.059 & 0.004 & 0.180 & 0.028 & 0.163 & 0.064 &  0.154 \\

Anxiety & -0.025 & 0.152 & -0.069 & 0.271 & 0.171 & 0.163 & 0.166 & 0.132 & 0.017 & 0.150 \\

Sadness & 0.283 & -0.089 & 0.152 & -0.032 & 0.168 & 0.021 & -0.051 & -0.059 & 0.258 & 0.008 \\

\toprule
\end{tabular}
\end{table*}

\begin{table*}[]
\setlength\tabcolsep{1.5pt}
\centering
\caption{Classification results for moral foundations using ZSC+SVM. F-1 is used to evaluate the model performance. }\label{tabreO21dErSviP} 
\begin{tabular}{lcccccccccc} \toprule
            & Care & Harm & Fairness & Cheating & Loyalty & Betrayal & Authority & Subversion & Purity & Degradation \\ \toprule
Lockdown & 0.671 & 0.673 & 0.704 & 0.647 & 0.688 & 0.708 & 0.706 & 0.710 & 0.640 & 0.692 \\
Wexit & 0.697 & 0.630 & 0.708 & 0.669 & 0.697 & 0.658 & 0.671 & 0.693 & 0.688 & 0.645 \\ \toprule
\end{tabular} 
\end{table*}

Research reveals that many Canadians have seen their stress levels double since the onset of the pandemic and are struggling with fear and uncertainty about their own and their loved ones' health \cite{CAMH2020}. Survey research conducted by Mental Health Research Canada found that feelings of depression are rising constantly \cite{mhrc2020}. Before the pandemic, 7\% of Canadians reported high levels of depression. This rate has risen to 16\% during the lockdown period and 22\% predict high levels of depression if social isolation continues for two more months. Our results are alarming and indicate potential signals relevant to mental health that can aid mental health services in assessing the impact of the pandemic on the population and implementing healthier coping strategies to build resilience.

{As for the WEXIT, it refers to a movement that advocates for separation from Canada. We report that the correlations between all vice moral foundations and {\tt negative emotion} and {\tt anger} as well as all virtue moral foundations and {\tt anxiety} and {\tt sadness} are statistically significant at $p<0.001$. We also note that there are some significant correlations between vice moral foundations and {\tt negative emotion}, and virtue moral foundations and {\tt positive emotion} ($p<0.05$). Our results show relatively low-levels of {\tt sadness} for {\tt Harm}, {\tt Subversion}, and {\tt Degradation} ($p>0.05$). We observe that WEXIT conversations include dominant emotional traits for vice moral foundations. This could be considered as strong evidence to argue that this trending topic may have sparked stormy debates and emotional statements. Identifying WEXIT proponents and opponents normally requires further analysis such as stance detection \cite{tshimula2020a}, measures of divergent opinions, and community detection to track persistent members of ever-growing communities and their linguistic idiosyncrasies. 







}


Table \ref{tabreO21dErSviP} presents the performance results for virtue and vice moral dimensions classification during coronavirus lockdown measures and the WEXIT movement in Canada. We observe that the F-1 scores are higher and show the ability to predict moral foundations, with F-1 scores of over 0.6 for the ten classes. Our model leverages large-scale social media text data stemming from trending topics. The absence of adequate annotated data on moral foundations may be challenging. To overcome this problem, we used psychologically validated and annotated dictionaries that indicate moral foundations' dimensions. We applied these resources to track moral foundations using ZSC, a model that does not require data to be annotated beforehand to predict text data; it learns a classifier on one set of labels and then evaluates on a different set of labels that the classifier has never seen before. Note that the lack of annotated data does not affect the generalizability of the findings and model performance.

\section{Summary and ongoing work}
We presented the first experiments towards investigating moral foundations from large-scale social media text data from trending topics. Our results provide strong evidence that we can predict moral foundations with an accuracy exceeding 0.6 and we can jointly investigate emotional traits and moral foundations. Though the results are encouraging, this problem leaves room for future research. In the future, we aim to study natural language inference and behavioral deterioration in moral foundations \cite{tshimula2020}. 

\bibliographystyle{unsrt}

\end{document}